\documentclass[
aip,jcp,amsmath,amssymb,reprint
]{revtex4-1}

\usepackage{graphicx}
\usepackage{dcolumn}
\usepackage{bm}

\begin{document}

\title[Structural flexibility of IDPs induces stepwise target recognition]
{Structural flexibility of intrinsically disordered proteins induces stepwise target recognition}

\author{Nobu C. Shirai}
\email{shirai@cp.cmc.osaka-u.ac.jp}
\affiliation{Graduate School of Science, Osaka University, Toyonaka, Osaka 560-0043, Japan}
\affiliation{Cybermedia Center, Osaka University, Toyonaka, Osaka 560-0043, Japan}

\author{Macoto Kikuchi}
\affiliation{Cybermedia Center, Osaka University, Toyonaka, Osaka 560-0043, Japan}
\affiliation{Graduate School of Science, Osaka University, Toyonaka, Osaka 560-0043, Japan}
\affiliation{Graduate School of Frontier Biosciences, Osaka University, Suita, Osaka 565-0871, Japan}

\date{\today}

\begin{abstract}
An intrinsically disordered protein (IDP) lacks a stable three-dimensional structure, while it folds into a specific structure when it binds to a target molecule.
In some IDP-target complexes, not all target binding surfaces are exposed on the outside, and intermediate states are observed in their binding processes. 
We consider that stepwise target recognition via intermediate states is a characteristic of IDP binding to targets with ``hidden" binding sites. 
To investigate IDP binding to hidden target binding sites, we constructed an IDP lattice model based on the HP model. 
In our model, the IDP is modeled as a chain and the target is modeled as a highly coarse-grained object. 
We introduced motion and internal interactions to the target to hide its binding sites. 
In the case of unhidden binding sites, a two-state transition between the free states and a bound state is observed, and we consider that this represents coupled folding and binding. 
Introducing hidden binding sites, we found an intermediate bound state in which the IDP forms various structures to temporarily stabilize the complex. 
The intermediate state provides a scaffold for the IDP to access the hidden binding site. 
We call this process multiform binding. 
We conclude that structural flexibility of IDPs enables them to access hidden binding sites, and this is a functional advantage of IDPs. 
\end{abstract}

\maketitle

\section{Introduction}
Intrinsically disordered proteins (IDPs) are biologically functional proteins without tertiary structure~\cite{DysonWright2005,TompaBook2009,DunkerObradovic2001}. 
It has been predicted that one-third of eukaryotic proteins are IDPs\cite{WardJones2004,XueUversky2012,YanKurgan2013}.
Many of them have specific binding partners, and characteristic binding processes called coupled folding and binding have been observed with nuclear magnetic resonance (NMR)~\cite{WrightDyson1999}. 
In these processes, IDPs bind to their partners and fold into a specific structure~\cite{KussiePavletich1996,UesugiVerdine1997,RadhakrishnanWright1997,SugaseWright2007,PapoianWolynes2003}. 
Binding to nucleic acids leads to transcriptional or translational regulation, and binding to a signaling target leads to signal transduction. 
Many IDPs are known as hub proteins in signaling networks~\cite{DunkerUversky2005,HaynesIakoucheva2006,HigurashiKinoshita2008,PatilNakamura2010}, and the evolutionary persistence of IDPs under selective pressure makes us think of functional advantages over globular proteins in signaling processes. 
One of the proposed advantages of IDPs is adaptability of their shapes to several binding targets, which is a suitable property for signaling hubs~\cite{KriwackiWright1996,Tompa2002}. 
Another proposed advantage of IDPs is the fast binding process due to its comparatively long capture radius~\cite{ShoemakerWolynes2000}, which is called the fly-casting mechanism. 
In this study, we propose a new advantage of IDPs, focusing on how they recognize their targets. 

In the recognition process, an IDP transiently interacts with a tentative target and searches for a biologically functional bound form. 
If the transient complex is not sufficiently stabilized by the intermolecular interaction, the complex is finally dissociated by thermal fluctuations with a comparatively short lifetime. 
Since the lifetime depends on the free-energy barrier to dissociation, stabilization of the complex in the early stage of the binding process is important to successfully recognize the correct target at an accidental encounter. 

In some of IDP-target complexes, however, not all of the interaction sites are exposed outside of the target molecule and easily accessible, and the p27$^\text{Kip1}$/cyclin A/cyclin-dependent kinase 2 (Cdk2) complex is a good example of this. 
p27 is known to be an IDP~\cite{KriwackiWright1996,LacyKriwacki2004}, and binds to the binary complex of cyclin A and Cdk2~\cite{ToyoshimaHunter1994}. 
Cdk2 of the binary complex has intramolecular hydrophobic interaction sites~\cite{JeffreyPavletich1995,RussoPavletich1996natStruct} which are finally exposed to p27 upon binding~\cite{RussoPavletich1996nature}. 
As another example, the phosphorylated kinase inducible activation domain (pKID) of the transcription factor cyclic-AMP-response-element-binding protein (CREB) , which is known to be an IDP~\cite{RadhakrishnanWright1997}, binds to the KID-binding (KID) domain of CREB binding protein inserting one of the hydrophobic residues deeply into the buried interaction pocket of KIX~\cite{RadhakrishnanWright1997}. 
In both examples with hidden binding sites, binding processes are not simple two-state transitions between a dissociated state and a bound state. 
It has been shown using isothermal titration calorimetry (ITC) and surface plasmon resonance (SPR) that p27 binds cyclin A before it binds to Cdk2~\cite{LacyKriwacki2004}. 
In the binding process between pKID and KIX, the intermediate bound state has been observed by NMR titrations and $^\text{15}$N relaxation dispersion~\cite{SugaseWright2007}. 
In this intermediate state, the buried interaction site of the target does not completely interact with pKID. 
We consider that the stepwise target recognition is a characteristic binding process between an IDP and its target with ``hidden" binding sites and a functional advantage of IDPs enable IDPs to access these hidden binding sites. 
In this paper, we construct a lattice model of an IDP based on the extended HP model~\cite{LauDill1989}, and investigate the target recognition processes using computer simulations of this model. 

\section{Model and method} \label{sec:model}
\begin{figure}
\begin{center}
\includegraphics[width=0.4\textwidth]{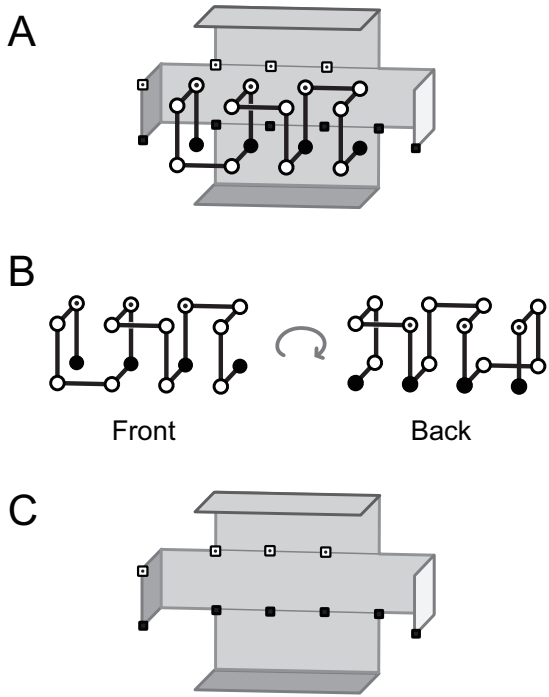}
\end{center}
\caption{
(A) The designed native bound structure is composed of the IDP which forms a helical structure and the binding surface of the target molecule. 
The IDP is modeled as a chain and the target is modeled as a combination of plates which represent the binding surface. 
{\it H}, {\it H$^\prime$} and {\it P} residues of the IDP are denoted by filled circles, circles with a dot, and open circles, respectively. 
Filled squares on the binding surface denote {\it H} interaction sites and squares with a dot denote {\it H$^\prime$} interaction sites. 
(B) Hydrophobic residues come together on one side of the helical structure of the IDP to form an amphipathic helix. 
(C) Target conformation in the native bound state. 
}
\label{fig:Fig1}
\end{figure}
In the HP model, proteins are simplified as two-letter sequences of polar ({\it P}) and hydrophobic ({\it H}) residues and only hydrophobic interactions between {\it H} residues are considered. 
This model has been used to analyze the kinetics and thermodynamics of protein folding~\cite{CamachoThirumalai1993,DillChan1997,ChikenjiKikuchi2000}, in which the hydrophobic interactions play a major role. 
Although IDPs themselves do not contain sufficient hydrophobic residues to stabilize a specific structure~\cite{DunkerObradovic2001}, hydrophobic interactions are still important for IDPs to bind to their target molecules~\cite{KussiePavletich1996,RussoPavletich1996nature,UesugiVerdine1997,RadhakrishnanWright1997,NomuraNishimura2005}. 
We then considered that the HP model can be applied to a binding process of an IDP and a target. 

We assumed two important properties which an IDP and its target should have. 
Firstly, we assumed that IDPs do not form specific structures in free states. 
Secondly, we assumed that the IDP and its target have a native bound state which is a non-degenerate ground state, and that the IDP adopts a specific structure in this bound state. 
The second assumption comes from the fact that many of the specific bound forms of IDP-target complexes were detected by crystal or NMR structures with sufficient stability. 
We introduce the second type of hydrophobic residues ({\it H$^\prime$}) to the HP model. 
In this extended model, a protein is modeled as a sequence of three types of residues: two types of hydrophobic residues ({\it H}, {\it H$^\prime$}) and one type of polar residue ({\it P}). 
We constructed a model of IDP by using this model, in which the IDP is modeled as a chain and the target is modeled as a highly coarse-grained object designed as a combination of plates which represent the binding surface. 
We designed a unique ground state of these two molecules as a bound state of an IDP and its target. 
The designed bound state is shown in Fig.~\ref{fig:Fig1}A, and we call it the native bound state in this paper. 
By separating the IDP and the target of this bound state, we get Fig.~\ref{fig:Fig1}B (IDP) and Fig.~\ref{fig:Fig1}C (target). 
In the native bound state, the IDP forms an amphipathic helix, while it lacks a specific structure in the ground state of the free states. 
\begin{figure}[h]
\includegraphics[width=0.4\textwidth]{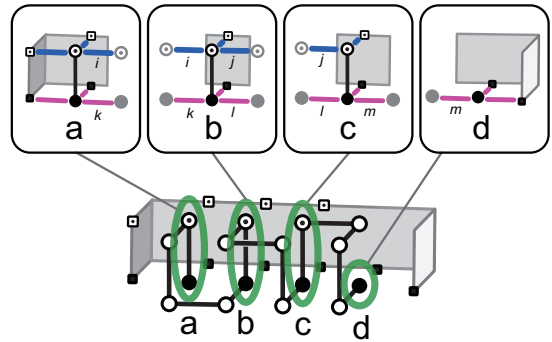}
\caption{
\label{fig:Fig2}
Hydrophobic interactions in the native bound state. 
The hydrophobic residues of the helical structure are divided into four parts (a-d), and interactions between hydrophobic residues are colored pink ({\it H}-{\it H} interactions) and blue ({\it H$^\prime$}-{\it H$^\prime$} interactions). 
The internal interactions of the IDP are labeled {\it i}-{\it m}.
}
\end{figure}

Each residue of an IDP is at a lattice point, and the whole protein is represented as a self-avoiding chain. 
We assumed that there are three types of interactions between (i) {\it H}-{\it H}, (ii) {\it H$^\prime$}-{\it H$^\prime$} and (iii) {\it H}-{\it H$^\prime$} pairs at nearest-neighboring sites, except pairs of consecutive residues along the chain. 
Contact energies of these interactions are denoted by $\varepsilon_{HH}$, $\varepsilon_{H^{\prime}H^{\prime}}$ and $\varepsilon_{HH^{\prime}}$ ($\varepsilon_{HH}, \varepsilon_{H^{\prime}H^{\prime}}, \varepsilon_{HH^{\prime}} < 0$), respectively. 
It should be noted that the hydrophobic interactions include solvent entropy, and thus ``contact energy" actually means ``contact free energy". 
We use $(\varepsilon_{HH}, \varepsilon_{H^{\prime}H^{\prime}}, \varepsilon_{HH^{\prime}}) = (-2,-1,-1)$ throughout this paper. 
We assumed that there are no interactions between {\it H}-{\it P}, {\it H$^\prime$}-{\it P} and {\it P}-{\it P} pairs. 
Using the extended HP model, we designed a 16-residue sequence of an IDP as {\it HH$^\prime$P$_3$HH$^\prime$P$_3$HH$^\prime$P$_3$H}. 
In this sequence, the hydrophobic residues come together on one side of the helical structure to form an amphipathic helix (Fig.~\ref{fig:Fig1}B). 
In the helical structure, there are three {\it H-H} interactions ({\it k}, {\it l} and {\it m} in Fig.~\ref{fig:Fig2}) and two {\it H$^\prime$-H$^\prime$} interactions ({\it i} and {\it j} in Fig.~\ref{fig:Fig2}), and the total energy of the intramolecular interactions of the IDP ($E_\text{ID}$) is $-8$. 
Many different conformations also have the same energy. 
The ground state of the IDP has energy of $-10$ and is forty eight-fold degenerate. 
Two of the ground state conformations are shown in Fig.~\ref{fig:Fig4}B as examples. 
\begin{figure}
\centering
\includegraphics[width=0.45\textwidth]{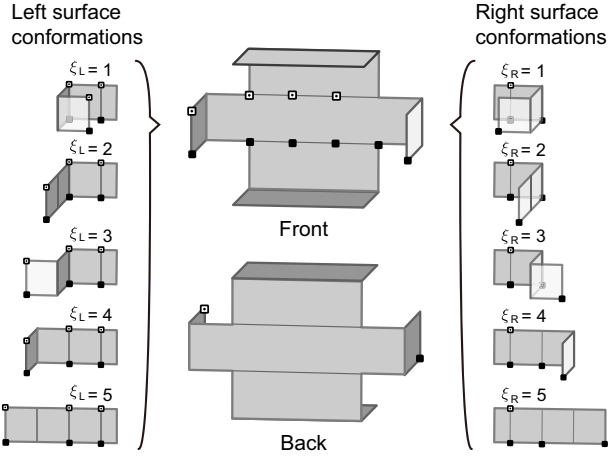}
\caption{
\label{fig:Fig3}
Conformations of the left and right parts of the designed target. 
The left part has five conformations labeled by $\xi_\text{L}$ ($\xi_\text{L} \in \{1,\cdots ,5\}$) and the right part has five conformations labeled by $\xi_\text{R}$ ($\xi_\text{R} \in \{1,\cdots ,5\}$). 
There are no interaction sites on the back of the binding surface. 
}
\end{figure}

We modeled only the binding surface of the target in order to focus on the recognition processes. 
Combining square plates and placing {\it H} and {\it H$^\prime$} interaction sites on them, we designed the binding surface of the native bound state shown in Fig.~\ref{fig:Fig1}C. 
In this conformation, all the interaction sites are exposed. 
Interactions between the IDP and the interaction sites of the binding surface are also given by the contact energies $\varepsilon_{HH}$, $\varepsilon_{H^{\prime}H^{\prime}}$ and $\varepsilon_{HH^{\prime}}$. 
The IDP cannot share the same site with the binding surface due to the excluded volume effect. 
As shown in Fig.~\ref{fig:Fig2}, there are six {\it H}-{\it H} interactions and four {\it H$^\prime$}-{\it H$^\prime$} interactions between the IDP and the binding surface in the native bound state, and the total energy of the intermolecular interactions ($E_\text{ID-T}$) is $-16$. 
Adding $E_\text{ID}$ to $E_\text{ID-T}$ of the native bound state, we get the ground state energy $E=-24$. 

In order to express hidden binding sites of the target, we remodeled the designed binding surface. 
Starting from the exposed binding surface shown in Fig.~\ref{fig:Fig1}C, we introduced motion of both sides of the target to open or close the binding surface by adding five conformations for each side, which are shown in Fig.~\ref{fig:Fig3}. 
The left and right sides of the conformations are labeled by $\xi_\text{L}$ $(\xi_\text{L} \in \{1,\cdots ,5\})$ and $\xi_\text{R}$ $(\xi_\text{R} \in \{1,\cdots ,5\})$, respectively, and a conformation of the whole binding surface is denoted by $(\xi_\text{L},\xi_\text{R})$. 
An example is the conformation of the binding surface in the native bound state shown in Fig.~\ref{fig:Fig1}C is $(4,4)$. 
In the conformations of $\xi_\text{L}=4 \text{ or } 5$ (or $\xi_\text{R}=4 \text{ and } 5$), all of the interaction sites are exposed outside, and in the other conformations, some of them are hidden. 
\begin{table}
\caption{
\label{table:RFHdefinition}
Definitions of surface types for each side of the binding surface. 
Conformations of the R surface are limited to $\xi_\text{L}=4$ (or $\xi_\text{R}=4$), while those of the F and H surfaces take $\xi_\text{L}=1-5$ (or $\xi_\text{R}=1-5$). 
Only the H surface has the internal interactions to close the binding surface, which come from non-zero $\lambda_\text{L}$ (or $\lambda_\text{R}$), while the R and F surfaces do not have internal interactions. 
}
\begin{ruledtabular}
\begin{tabular}{lccclcc}
Surface type & $\xi_\text{L}$ & $\lambda_\text{L}$ & &
Surface type & $\xi_\text{R}$ & $\lambda_\text{R}$ \\
\hline
R (rigid) & $4$ & $0$ & & R (rigid) & $4$ & $0$ \\
F (flexible) & $1-5$ & $0$ & & F (flexible) & $1-5$ & $0$ \\
H (hidden) & $1-5$ & $1$ & & H (hidden) & $1-5$ & $1$ \\
\end{tabular}
\end{ruledtabular}
\end{table}
To energetically stabilize a conformation with hidden binding sites, we introduced intramolecular interactions between the interaction sites of the target by defining the energy of the target as
\begin{equation}
E_\text{T} = 
\lambda_\text{L}
( \varepsilon_{HH} +
\varepsilon_{H^{\prime}H^{\prime}} )
\delta_{\xi_\text{L}1}
+
\lambda_\text{R}
\varepsilon_{HH}
\delta_{\xi_\text{R}1},
\end{equation}
where $\lambda_\text{L}$ and $\lambda_\text{R}$ are parameters which take 0 or 1 to control the intramolecular interactions. 
By setting $\lambda_\text{L}=1$ (or $\lambda_\text{R}=1$) , we can introduce hidden binding sites. 
$\delta_{ij}$ is Kronecker's delta given by
\[
\delta_{ij} =
\begin{cases}
1, & \text{if } i=j, \\
0, & \text{otherwise}.
\end{cases}
\]

Using the conformations and the intramolecular interactions, we defined the three surface types, which are denoted by R (rigid), F (flexible) and H (hidden), for each side of the binding surface. 
The defined surface types are shown in Table~\ref{table:RFHdefinition}. 
The R surface is fixed in the form of $\xi_\text{L}=4$ (or $\xi_\text{R}=4$), and the binding sites are always exposed. 
The F surface has flexibility to change its conformation $\xi_\text{L}$ (or $\xi_\text{R}$) from $1$ to $5$. 
All of these conformations have the same energy. 
The H surface is also flexible, while the hidden conformation $\xi_\text{L}=1$ (or $\xi_\text{R}=1$) is stabilized by the internal interactions which come from non-zero $\lambda_\text{L}$ (or $\lambda_\text{R})$. 
By selecting the surface types of both sides, we define the whole surface type of the target and denote it by two-letter prefix, such as ``HF-target", which means that the left surface is hidden and the right surface is flexible. 

We confined the IDP and the target in an $L\times L \times L$ cubic lattice space surrounded by walls. 
The target is fixed at the center of the system, while the IDP can move around inside the system. 
The total energy of the system is given by
\begin{equation}
E=E_\text{ID} + E_\text{T} + E_\text{ID-T},
\end{equation}
and the partition function is written as
\begin{equation}
Z=\sum_{E} g(E) {\mathrm e}^{-E/T},
\end{equation}
where $g(E)$ is the number of states with energy $E$. 
$T$ is temperature (we take the Boltzmann constant $k_B=1$). 
We also define the numbers of states $g(A,E)$ and $g(A,B,E)$ for the physical quantities $A$ and $B$. 
Further, $g(A,E)$ and $g(A,B,E)$ fulfill
\begin{equation}
\sum_A g(A,E) = \sum_A \sum_B g(A,B,E) = g(E).
\end{equation}
Using $g(A,E)$ and $g(A,B,E)$, we define free energies with one and two variables by
\begin{equation} \label{eq:FreeErgA}
F(A) = -T \log \left\{ \sum_E g(A,E) \mathrm{e}^{-E/T} \right\} + F_0,
\end{equation}
\begin{equation} \label{eq:FreeErgAB}
F(A,B) = -T \log \left\{ \sum_E g(A,B,E) \mathrm{e}^{-E/T} \right\} + F_0,
\end{equation}
where $F_0$ is a constant reference value of the free energies. 

We calculated the number of states by using the multi-self-overlap ensemble (MSOE) Monte Carlo method~\cite{IbaKikuchi1998,ChikenjiIba1999}, which is an extended version of the multicanonical Monte Carlo method~\cite{BergNeuhaus1991,BergNeuhaus1992}. 
The MSOE is a powerful tool in analyzing thermal properties of lattice polymers~\cite{ChikenjiKikuchi2000,NakanishiKikuchi2006}, and is applied for statistical enumeration of self-avoiding walks~\cite{ShiraiKikuchi2013}. 
The MSOE successfully attained thermal equilibrium between free states and the designed bound state. 
More details of the method are given in Appendix~\ref{sec:calcFreeErgLand}. 

\section{Result and Discussion} \label{sec:result}
\subsection{Intrinsic disorder represented as a mixture of several states}
\begin{figure}
\includegraphics[width=0.45\textwidth]{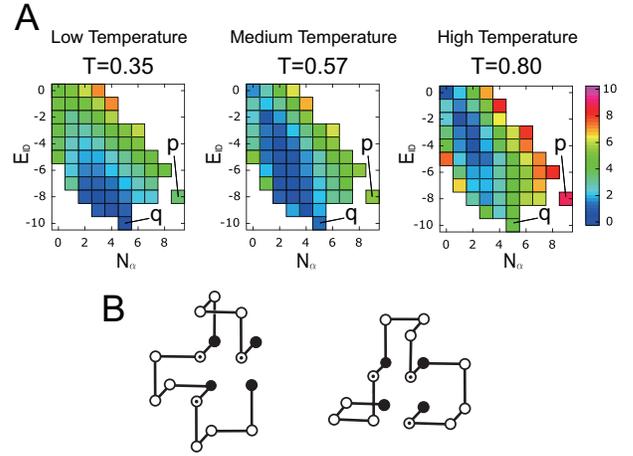}
\caption{
\label{fig:Fig4}
(A) Free-energy landscapes $F(N_\alpha, E_\text{ID})$ of the IDP for $L=\infty$ at low, medium and high temperatures shown from left to right, respectively. 
The medium temperature is the peak temperature of the specific heat of the IDP for $L=\infty$ ($T_\text{ID}$) shown in Fig.~\ref{fig:Fig13}. 
Colors indicate the values of free energy.
(B) Two example conformations of the ground state (q). 
}
\end{figure}
\begin{figure}
\includegraphics[width=0.4\textwidth]{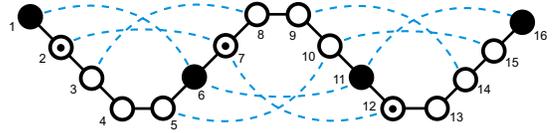}
\caption{
\label{fig:Fig5}
The $\alpha$-helix contact pairs are defined by the residue pairs connected by blue dotted lines. 
The definition of the $\alpha$-helix contact $N_\alpha$ is the number of the $\alpha$-helix contact pairs which are at nearest-neighbor sites. 
In the helical structure shown in Fig.~\ref{fig:Fig1}B and its mirror image, all these contact pairs are at nearest-neighbor sites and $N_\alpha$ takes the maximum value of $9$. 
}
\end{figure}

Free-energy landscapes $F(N_\alpha, E_\text{ID})$ of the IDP at three temperatures are shown in Fig.~\ref{fig:Fig4}A. 
$N_\alpha$ shown on the $x$-axis indicates structural similarity to a helical structure, and the definition of $N_\alpha$ is shown in Fig.~\ref{fig:Fig5}. 
At the high temperature, the IDP is unfolded and extended, and it is denatured. 
As temperature decreases ($T=0.80 \to 0.57 \to 0.35$), the distribution of the low free-energy states (shown as blue areas in Fig.~\ref{fig:Fig4}A) moves from high $E_\text{ID}$ states to low $E_\text{ID}$ states. 
The states at point (p) in Fig.~\ref{fig:Fig4}A, which includes the helical structure shown in Fig.~\ref{fig:Fig1}B and its mirror image, are always less favorable.  In this temperature range, the IDP is in a mixture of many states and it does not stay in a specific structure. 
Below these temperatures, the blue area of the landscape converges to the ground state of the IDP at the point (q) of Fig.~\ref{fig:Fig4}A, which is a glassy state of many collapsed structures. 
Two of the ground state conformations of the IDP are shown in Fig.~\ref{fig:Fig4}B as examples. 
We confirmed that the constructed model actually has the properties shown in the second paragraph of Sec.~\ref{sec:model} and we successfully constructed a model of an IDP. 
$L$-dependence of the specific heat of the IDP is shown in Appendix~\ref{sec:SizeDependence}. It shows that the peak temperature of the specific heat does not strongly depend on $L$. 

\subsection{Thermal stability of the native bound state}
\begin{figure}
\includegraphics[width=0.45\textwidth]{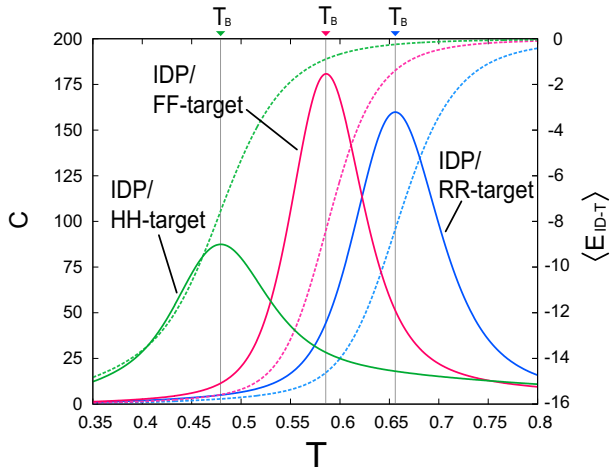}
\caption{
\label{fig:SpecificHeatL32}
The specific heat ($C$) for $L=32$ with an IDP and the three types of binding surfaces: RR-target (blue line), FF-target (pink line) and HH-target (green line). 
The binding temperature $T_\text{B}$ of each system is indicated by the vertical line. 
We also show the thermal average of the contact energy between the IDP and the target ($\langle E_\text{ID-T} \rangle$) of the same systems (broken lines in the same colors as the specific heats). 
}
\end{figure}

Fig.~\ref{fig:SpecificHeatL32} shows the specific heat for $L=32$ with the IDP and three types of the targets: RR-target, FF-target and HH-target. 
As temperature decreases, the change of $\langle E_\text{ID-T} \rangle$ occurs at the peak temperature of each system. 
At this temperature, the IDP binds to or dissociates from the target, and we call it binding temperature $T_\text{B}$. 
$T_\text{B}$ reflects the thermal stability of the native bound state. 
Comparing $T_\text{B}$ of the IDP/RR-target and IDP/FF-target systems, we find that the motion of the binding surface destabilizes the native bound state, and comparing $T_\text{B}$ of the IDP/FF-target and IDP/HH-target systems, we find that the native bound state is destabilized by the hidden binding sites. 
These results suggest that a change in surface types can induce binding of the IDP. 
For example, if post-translational modification or ligand binding causes a loss of degrees of freedom of the target, $T_\text{B}$ increases. 
If $T_\text{B}$ becomes larger than the cell temperature, the change of the target can induce binding of the IDP by enhancing the thermal stability of the bound state. 

\begin{figure}
\includegraphics[width=0.45\textwidth]{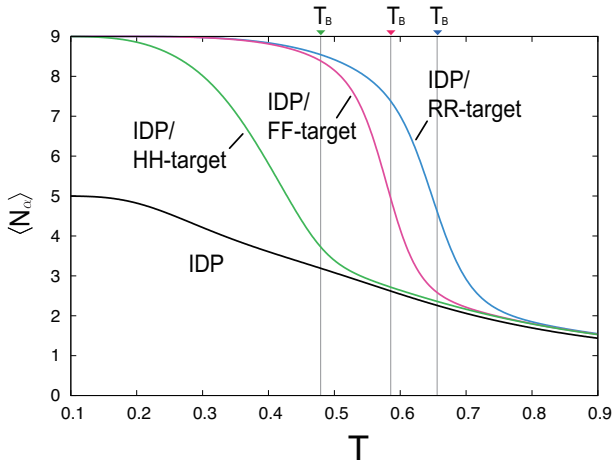}
\caption{
\label{fig:Fig7}
The thermal averages of the $\alpha$-helix contact ($\langle N_\alpha \rangle$) for the four systems ($L=32$): IDP only (black), IDP/RR-target (blue), IDP/FF-target (pink) and IDP/HH-target (green).
$T_\text{B}$ for the three IDP-target systems are indicated by the vertical lines. 
}
\end{figure}
In order to follow structural changes of the IDP at the binding temperatures, we show the thermal averages of $N_\alpha$ in Fig.~\ref{fig:Fig7}. 
As temperature decreases, $\langle N_\alpha \rangle$ of the IDP gradually converges to $N_\alpha=5$, which is the value of the ground state conformations. 
By adding each of the three types of the targets to the system, the curve $\langle N_\alpha \rangle$ reaches the maximum value of $9$ at the low temperature, and the IDP folds into the helical structure below $T_\text{B}$ of each system. 
\begin{figure}
\includegraphics[width=0.45\textwidth]{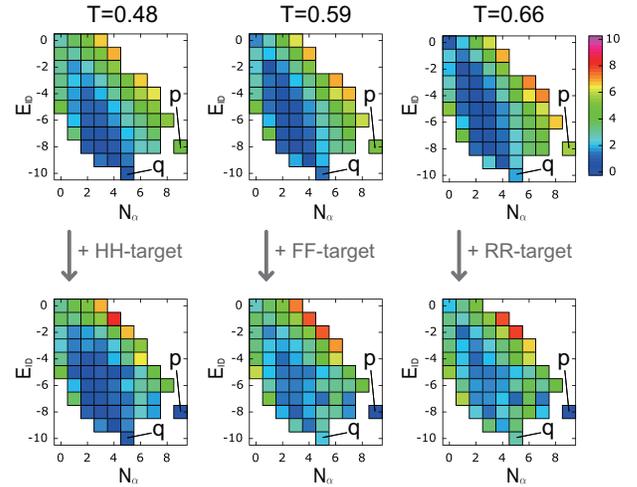}
\caption{
\label{fig:Fig8}
Free-energy landscapes $F(N_\alpha, E_\text{ID})$ of the IDP for $L=32$ with and without the three types of binding surfaces: RR-target (top and bottom left), FF-target (top and bottom center) and HH-target (top and bottom right) at the corresponding binding temperatures. 
}
\end{figure}
In Fig.~\ref{fig:Fig8}, we show how the free-energy landscapes $F(N_\alpha, E_\text{ID})$ of the IDP change according to the presence of the three types of binding surfaces at the corresponding binding temperatures. 
In all cases, the helical structure of the IDP is stabilized by the presence of the targets. 
Thus, we consider binding to the target induces folding processes and this is nothing but coupled folding and binding. 
This is consistent with the discussion by Matsushita and Kikuchi~\cite{MatsushitaKikuchi2013} in which the presence of a target reduces the inconsistency of conformations of an IDP. 
In the systems of the IDP/RR-target and IDP/FF-target, the change of the curves occurs just around $T_\text{B}$, and this suggests the coupled folding and binding in equilibrium induced by temperature change. 
The change of the curve of the IDP/HH-system occurs at a lower temperature than $T_\text{B}$, and this might be caused by multistep binding which we explain in Sec.~\ref{subsec:stepwise}. 

\begin{figure}[h]
\includegraphics[width=0.45\textwidth]{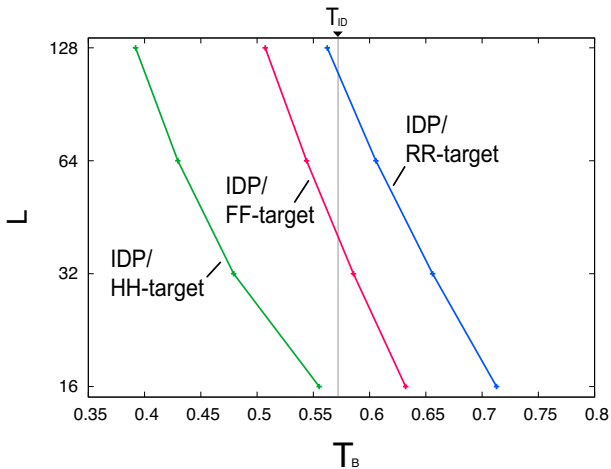}
\caption{
\label{fig:Fig9}
System size dependence ($L=16-128$) of the binding temperature $T_\text{B}$. 
$T_\text{ID}$, which is indicated by the vertical line, is the peak temperature of the specific heat of the IDP for $L=\infty$ shown in Fig.~\ref{fig:Fig13}. 
}
\end{figure}

The thermal stability of the complexes depends on the system size.
System size dependence of $T_\text{B}$ is shown in Fig.~\ref{fig:Fig9}. 
As $L$ decreases, the thermal stability of the complexes increases because of the decrease in the translational entropy of the free states. 
This result suggests that confinement in a small volume can also induce binding, which is basically the same as the confinement-induced dimerization discussed by Nakanishi and Kikuchi~\cite{NakanishiKikuchi2006}. 
Increasing the density of the two molecules can also reduce translational entropy of the free states, and stabilize bound states. 
It is known that IDPs of eukaryotes localize in the nucleus~\cite{WardJones2004}, and we consider that this localization stabilizes bound complexes of IDPs with nuclear proteins or DNAs by increasing their density. 
Macromolecular crowding also stabilizes bound states by reducing free volume. 

\begin{figure*}
\includegraphics[width=0.95\textwidth]{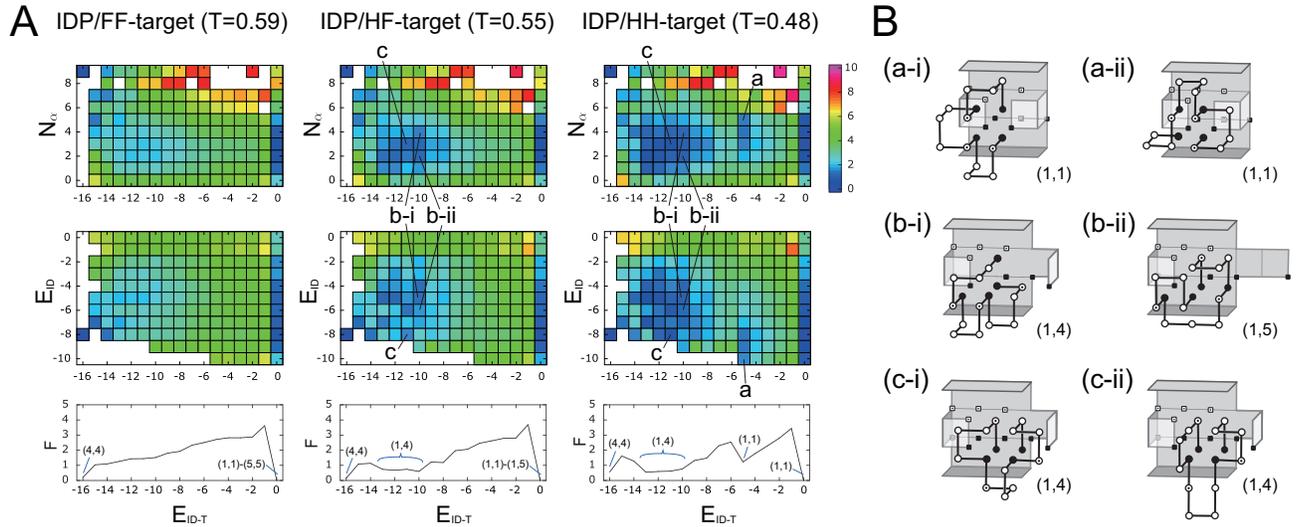}
\caption{
\label{fig:Fig10}
(A) Free-energy landscapes $F(E_\text{ID-T},E_\text{ID})$, $F(E_\text{ID-T},E_\text{N})$ and $F(E_\text{ID-T})$ for $L=32$ of IDP/FF-target, IDP/HF-target and IDP/HH-target at the binding temperature of each system. 
In the graphs of $F(E_\text{ID-T})$, we show the most stable conformations $(\xi_\text{L},\xi_\text{R})$ of the binding surface which is given in Figs.~\ref{fig:Fig12} and \ref{fig:Fig14}. 
Colors indicate the values of free energy. 
(B) Example structures of the IDP/target complexes at the points denoted by (a-c) shown in A. 
The target conformations $(\xi_\text{L},\xi_\text{R})$ is shown in the bottom right of each figure. 
}
\end{figure*}

\subsection{Stepwise target recognition of the IDP due to multiform binding effect}
\label{subsec:stepwise}
Fig.~\ref{fig:Fig10}A shows the free-energy landscapes of IDP/FF-target, IDP/HF-target and IDP/HH-target systems at the $T_\text{B}$ of each system. 
All of the landscapes of $F(E_\text{ID-T})$ (bottom row) have local minima at $E_\text{ID-T}=0 \text{ and } -16$, corresponding to the free states and the native bound state, respectively. 
These landscapes show that the closure of the left surface (FF$\to$HF) stabilizes the non-native intermediate state around $E_\text{ID-T}=-12$, and further introduction of a hidden binding site (HF$\to$HH) stabilizes another intermediate state around $E_\text{ID-T}=-5$. 
In these intermediate states, the IDP forms various structures to stabilize the encounter complexes, some of which are shown in Fig.~\ref{fig:Fig10}B. 
Various non-native bound complexes are also found in the detailed simulation of the N-terminal repressor domain of neural restrictive silencer factor (NRSF), which is known to be an IDP, and the paired amphipathic helix domain of mSin3 (target), using classical molecular dynamics in atomic resolution~\cite{HigoNakamura2011}. 
We considered that these intermediate complexes are related to the advantages of IDPs. 

In the first step to bind to the HH-target, the IDP interacts with the exposed binding sites on the outside of the target by forming low-energy compact structures, two of which are shown in Fig.~\ref{fig:Fig10}B (a) as examples ($-10 \le E_\text{ID} \le -8$). 
These intermediate structures provide the first scaffold to open one side of hidden binding sites. 
After opening the one side of the binding surfaces, the IDP interacts with the surface to stabilize the second scaffold to access the other hidden binding sites. 
Some of the complex structures of the second intermediate state are shown in Figs.~\ref{fig:Fig10}B (b) and (c) as examples. 
The IDP is in a mixture of various states ($-8 \le E_\text{ID} \le -3$) which can increase contacts with the unclosed interaction sites. 
Lastly, the IDP opens the hidden binding site and forms the native bound structure. 
Thus, binding to the HH-target has three steps via two intermediate states. 
In the case of the HF-target, the binding process starts with the second intermediate state shown above, and it involves two steps. 
These results suggest that closure of binding sites stabilizes intermediate states and this mechanism simply explains the existence of the intermediate states of p27/cyclin A/Cdk2 and pKID/KIX systems. 

\begin{figure}
\begin{center}
\includegraphics[width=0.4\textwidth]{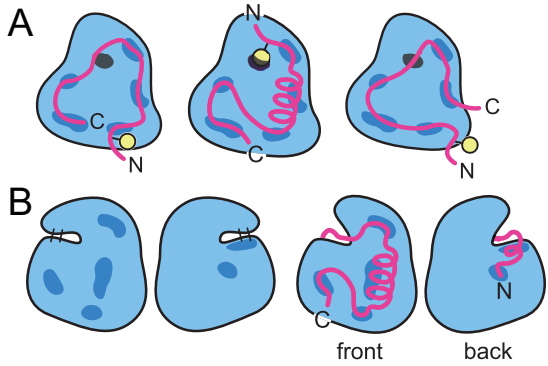}
\end{center}
\caption{
Conceptual pictures of multiform binding of two different hidden binding sites. 
(A) Several binding forms due to flexibility of an IDP (pink curves) stabilize an intermediate state with the target molecule (light blue objects with interaction sites shown as blue spots) and provide a scaffold to search for buried interaction sites (black spots) which interact with the hydrophobic residues of an IDP (yellow circle). 
(B) In other cases, several binding forms provide a scaffold to cut into intramolecular interactions of the target. 
The letters N and C shown in the figure indicate the N-terminus and C-terminus of IDPs, respectively. 
}
\label{fig:Fig11}
\end{figure}

These results give us a new insight into an advantage of IDPs shown in Fig.~\ref{fig:Fig11}. Various forms of an IDP in transient encounter complexes can interact with exposed binding sites on the outside of the target and can provide a scaffold to search for hidden binding sites (Figs.~\ref{fig:Fig11}A and \ref{fig:Fig11}B). 
We call this multiform binding. 
This is similar to the proposed advantage of structural adaptability of an IDP for several different targets. 
But in this case, we discuss the effect of multiple forms on binding to the same target. 

\subsection{Coupled folding and binding of the flexible target}
\begin{figure}
\includegraphics[width=0.48\textwidth]{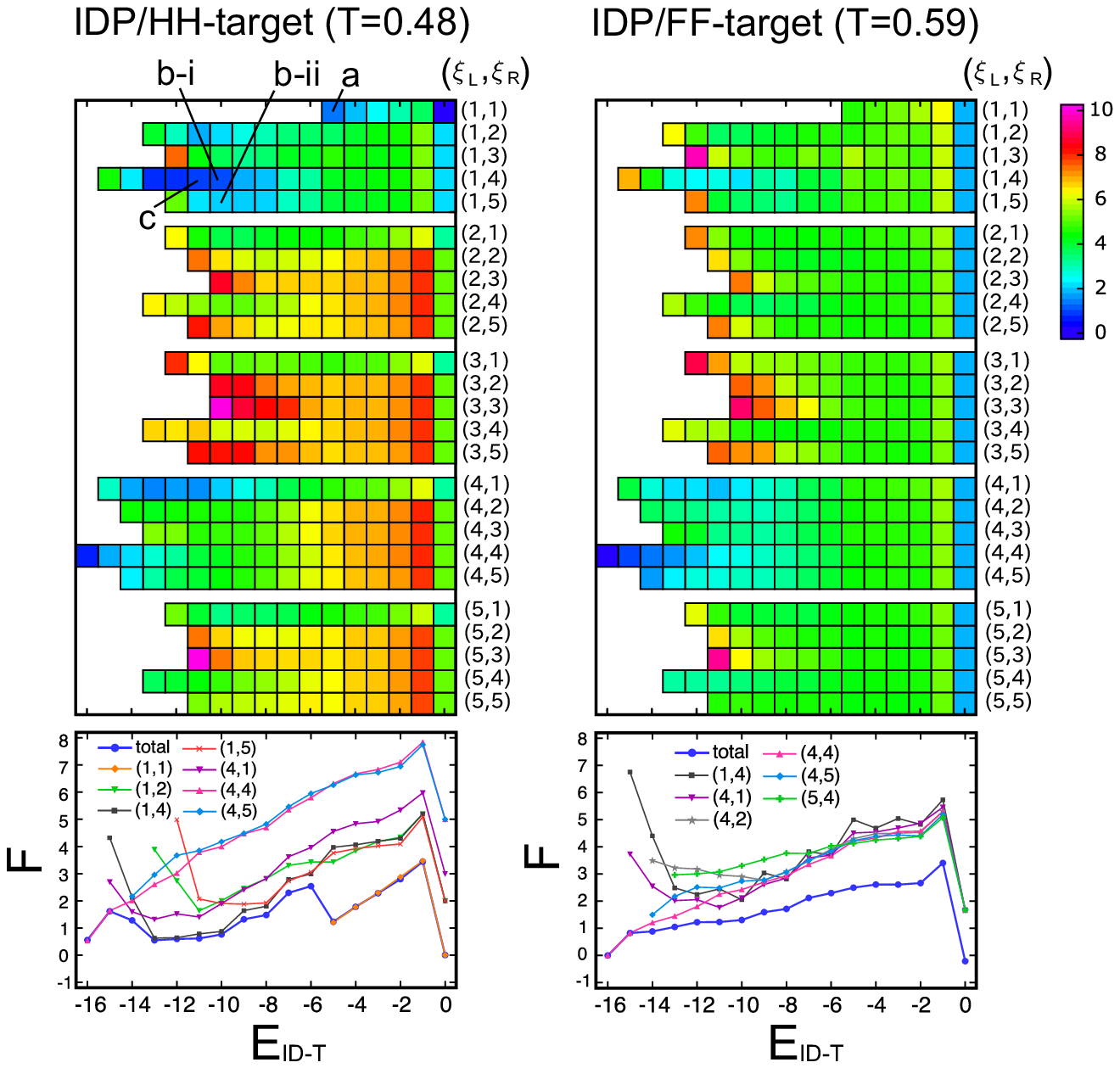}
\caption{
\label{fig:Fig12} 
Free-energy landscapes $F(E_\text{ID-T}, (\xi_\text{L},\xi_\text{R}))$ and $F(E_\text{ID-T})$ for $L=32$ of IDP/HH-target and IDP/FF-target at the binding temperature of each system. 
In the graph of $F(E_\text{ID-T})$ labeled ``total", we superposed lines of $F(E_\text{ID-T},(\xi_\text{L},\xi_\text{R}))$ for selected $(\xi_\text{L},\xi_\text{R})$ which greatly contribute to reducing the value of $F(E_\text{ID-T})$.
Colors indicate the values of free energy. 
The letters (b) and (c) correspond to the example structures of Fig.~\ref{fig:Fig11}B. 
}
\end{figure}
The blue lines labeled ``total" on the bottom-left and bottom-right of Fig.~\ref{fig:Fig12} show the free-energy landscapes $F(E_\text{ID-T})$ of the IDP/HH-target and IDP/FF-target systems, respectively. 
We obtain the other lines in the same figure by decomposing these graphs into $F(E_\text{ID-T}, (\xi_\text{L},\xi_\text{R}))$ which greatly contribute to reducing the value of $F(E_\text{ID-T})$. 
Using these landscapes, we can deduce the dynamics of the binding and folding processes.
Both sides of the HH-target are in the closed conformation $(1,1)$ in the free states, and the IDP makes contact with this conformation. 
After the target opens the right side of the hidden binding sites, conformations are sequentially selected as $\{(1,2),(1,4),(1,5)\}\to \{(1,4),(1,5)\} \to \{(1,4)\}$ before switching to the open conformation $(4,4)$. 
In this process, the conformational stabilization of the target is coupled with the binding process. 
This result suggests that the IDP and the target cooperatively reduce entropy of the complex and form a binding funnel. 
In the case of the IDP/FF-target system, all conformations of the target contribute to reducing the free-energy barrier to IDP binding, and after interacting with the IDP, the target reduces its conformational entropy and reaches $(4,4)$. 
There are no free-energy barriers to the open conformation $(4,4)$ in the final stage of the binding process. 
The IDP and the target also form a binding funnel in this process. 
We show two more examples of free-energy landscapes $F(E_\text{ID-T},(\xi_\text{L},\xi_\text{R}))$ in Appendix~\ref{sec:targetStateLandscape}. 

\section{Conclusion} \label{sec:discussion}
We represented an IDP as a mixture of various states, without a specific structure formed in an equilibrium state. 
The IDP binds to the target below its binding temperature and forms a helical structure. 
This represents coupled folding and binding. 
By comparing free-energy landscapes of various types of the binding surface, we found that closure of the binding surface produces a new intermediate state on the binding pathway and it can simply explain the existence of an intermediate state of p27/cyclin A/Cdk2 and pKID/KIX systems. 
From this result, we conclude that flexibility of the IDP provides a scaffold to access the closed binding site, which we call multiform binding, and this is a functional advantage of an IDP. 
By decomposing the free-energy landscape by the conformations of the target, we found the conformational selection process of the target. 
This result suggests that the IDP and the target form a binding funnel. 

\begin{acknowledgments}
This work was supported by a Grant-in-Aid (Grant No. 21113006) from MEXT Japan, and the Global COE Program Core Research and Engineering of Advanced Materials-Interdisciplinary Education Center for Materials Science from MEXT Japan. 
\end{acknowledgments}

\appendix
\section{Calculating free-energy landscapes by using the MSOE} \label{sec:calcFreeErgLand}
In the MSOE simulation, we relax the excluded volume condition, and allow the residues of the IDP to overlap with themselves and with the target. 
The MSOE makes it possible to explore the configuration space faster than with the conventional multicanonical ensemble method. 
Let $V$ be the number of overlaps. 
Introducing a weight $W(E,V)$ as a function of $E$ and $V$, we define the transition probability from a state of the system $s_a$ to another state $s_b$ as
\begin{equation}
p(s_a \to s_b) = 
\min\left[
\frac{W(E_b,V_b)}{W(E_a,V_a)},1
\right],
\end{equation}
where $E_a$ and $E_b$ are energy of $s_a$ and $s_b$, respectively. 
In the case of the IDP alone, we randomly select a residue and change the local conformation of the residue according to this transition probability. 
In the case of the system with both the IDP and the target, we first select the IDP or the target with a fixed probability, and then change the state of the selected object according to the transition probability. 
We iteratively improve $W(E,V)$ to be approximately proportional to the inverse of the number of states $1/g(E,V)$. 
Using the determined $W(E,V)$, the MSOE simulation produces a flat histogram $H(E,V)$. 
The $V=0$ subspace corresponds to the conventional multicanonical ensemble and the number of states $g(E)$ is given by $H(E,0)/W(E,0)$ up to a normalization factor. 
In this paper, we made the three- and four-dimensional histograms listed below to create free-energy landscapes. 
\begin{itemize}
\item $H(E_\text{ID},E,V)$ (for Fig.~\ref{fig:Fig10}A, Fig.~\ref{fig:Fig12} and Fig.~\ref{fig:Fig14})
\item $H(N_\alpha, E_\text{ID},E,V)$ (for Fig.~\ref{fig:Fig4}A and Fig.~\ref{fig:Fig8})
\item $H(E_\text{ID-T},N_\alpha,E,V)$, $H(E_\text{ID-T},E_\text{ID},E,V)$ (for Fig.~\ref{fig:Fig10}A)
\item $H(E_\text{ID-T},(\xi_\text{L},\xi_\text{R}),E,V)$, $H(E_\text{ID-T},(\xi_\text{L},\xi_\text{R}),E,V)$ (see $(\xi_\text{L},\xi_\text{R})$ as one variable; for Fig.~\ref{fig:Fig12} and Fig.~\ref{fig:Fig14})
\end{itemize}
From these histograms, we obtain free-energy landscapes using Eq.~\ref{eq:FreeErgA} and Eq.~\ref{eq:FreeErgAB}. 

\section{Size dependence of the IDP} \label{sec:SizeDependence}
\begin{figure}[h]
\includegraphics[width=0.45\textwidth]{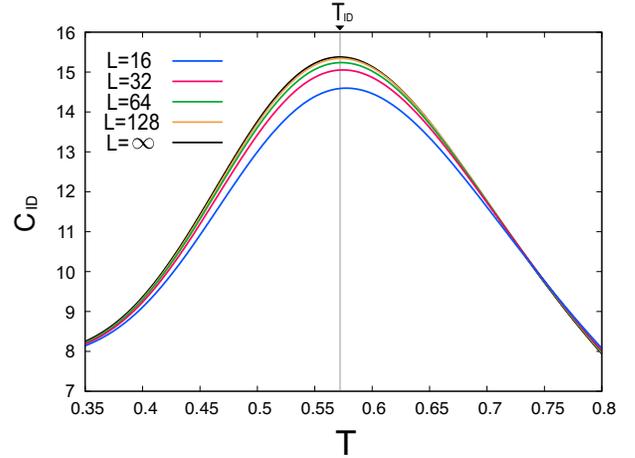}
\caption{
\label{fig:Fig13}
The specific heat of the IDP ($C_\text{ID}$) for $L=16, 32, 64, 128  \text{ and } \infty$. 
The peak temperature for $L=\infty$ ($T_\text{ID}$) is indicated by the vertical line. 
}
\end{figure}

The specific heat of the IDP ($C_\text{ID}$) for $L=16, 32, 64, 128 \text{ and } \infty$ is shown in Fig.~\ref{fig:Fig13}. 
We denote the peak temperature of $C_\text{ID}$ for $L=\infty$ by $T_\text{ID}$. 
As $L$ increases, the curves converge to the curve of $L=\infty$. 
The shapes and $T_\text{ID}$ of these curves do not strongly depend on the system size. 

\section{Other examples of the free-energy landscapes $F(E_\text{ID-T},(\xi_\text{L},\xi_\text{R}))$}
\label{sec:targetStateLandscape}
\begin{figure}
\includegraphics[width=0.48\textwidth]{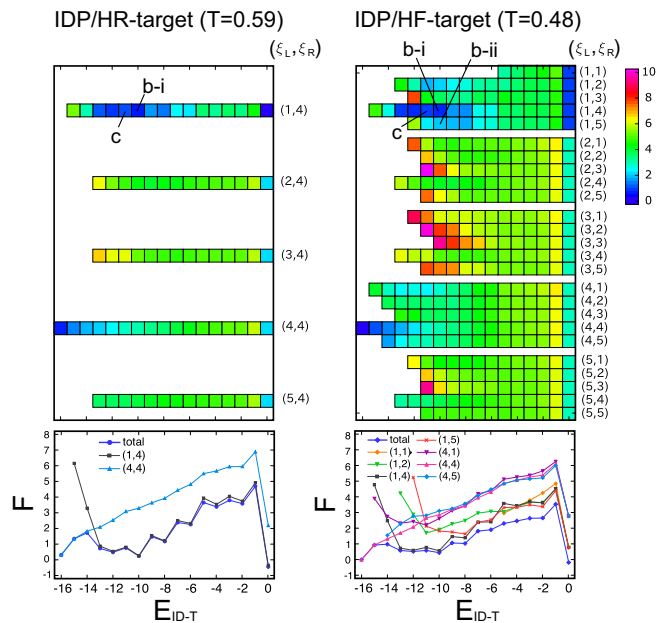}
\caption{
\label{fig:Fig14}
Free-energy landscapes $F(E_\text{ID-T}, (\xi_\text{L},\xi_\text{R}))$ and $F(E_\text{ID-T})$ for $L=32$ of IDP/HR-target and IDP/HF-target at the binding temperature of each system. 
In the graph of $F(E_\text{ID-T})$ labeled ``total", we superposed lines of $F(E_\text{ID-T},(\xi_\text{L},\xi_\text{R}))$ for selected $(\xi_\text{L},\xi_\text{R})$ which greatly contribute to reducing the value of $F(E_\text{ID-T})$.
Colors indicate the values of free energy. 
The letters (b) and (c) correspond to the example structures of Fig.~\ref{fig:Fig11}B. 
}
\end{figure}

The blue lines labeled ``total" on the bottom-left and bottom-right of Fig.~\ref{fig:Fig14} show the free-energy landscapes $F(E_\text{ID-T})$ of the IDP/HR-target and IDP/HF-target systems, respectively. 
We obtain the other lines in the same figure by decomposing these graphs into $F(E_\text{ID-T}, (\xi_\text{L},\xi_\text{R}))$ which greatly contribute to reducing the value of $F(E_\text{ID-T})$. 
In the case of the IDP/HR-target system, we observe only one route to the native bound state by switching from the closed conformation $(1,4)$ to the open conformation $(4,4)$. 
In this process, the folding of the IDP is coupled with the binding of the IDP to the target. 
By introducing flexibility into the right side of the target (HR$\to$HF), we observe a conformational selection process of the target before switching to the open conformation $(4,4)$. 
As $E_\text{ID-T}$ decreases, the target starts from the closed conformations $\{(1,1)-(1,5)\}$ in the free states and they are sequentially selected as $\{(1,1),(1,2),(1,4),(1,5)\}\to \{(1,4),(1,5)\} \to \{(1,4)\}$. 
In this process, the folding of the IDP is also coupled with the binding process, and a flexible binding surface gradually leads to the native bound state by reducing fluctuation. 
This result suggests that the IDP and the target cooperatively reduce entropy of the complex and form a binding funnel. 

\end{document}